# The fantastic single-molecule techniques

Huang Tang,[a,b,1] Shuting Liu,[a,b,1] Chenyue Kang,[a,b] Xiang Wang,[a,b,c] Xi Zhang,[a,b] Kun Li,[a,b] Gege Duan,[a,b] Zheng Li,[d,e,2] Boyang Hua[a,b,2]

[a]School of Chemistry and Chemical Engineering, Nanjing University, Nanjing, Jiangsu 210023, P. R. China
[b]State Key Laboratory of Analytical Chemistry for Life Science, Nanjing University, Nanjing, Jiangsu 210023, P. R. China
[c]Shenzhen Grubbs Institute and Department of Chemistry, Southern University of Science and Technology, Shenzhen, Guangdong 518055, P. R. China
[d]Department of Materials Science and Engineering, College of Chemistry and Materials Science, Jinan University, Guangzhou 511443, P. R. China
[e]Institute for Advanced Study, Shenzhen University, Shenzhen, Guangdong 518060, P. R. China

[1]H.T., and S.L. contributed equally to this work.
[2]To whom correspondence may be addressed.

*Abstract*

In the past 40 years, single-molecule techniques have been rapidly developed and widely applied in numerous fields of biology researches, offering new insights that conventional biochemical assays cannot discover. In this review, to help fully appreciate the powerfulness of single-molecule methods, we systemically summarize the various advantages of performing biochemical assays at the single-molecule level. Inspired by these examples, we propose a new single-molecule polysome profiling technique, to demonstrate that this strategy is not limited to the few special "outliers". Finally, we point out a possibility in the future of unifying different biochemical assays on the platform of single-molecule microscopy, which will reduce the cost of instrumentation and inevitably promote the applicability and adoptability of new biochemical and biophysical methods.

*Introduction*

The modern day single-molecule techniques encompass a huge family of methods and tools that can be loosely divided into three broad categories, *i.e.*, single-molecule electrical techniques (such as patch-clamps,[1] nanopores,[2, 3] and scanning tunneling microscopy[4]), optical techniques (such as fluorescence,[5, 6] Raman,[7, 8] and infrared spectroscopy[9-11]), and force techniques (such as atomic force microscopy,[12-15] optical tweezers,[16-20] and magnetic tweezers[21, 22]). It is a star-studded collection that bears not one but several Nobel prize-winning inventions. Although these techniques are built upon vastly different instrument principles and measure very different physical properties, the one thing they have in common is that they all carry the modifier "single-molecule". In a sense, any technique can be a "single-molecule" technique if it measures signals from individual molecules instead of from their ensemble. This begs the question, why "single-molecule"? The short answer to this question often goes as the following, unlike bulk measurements, single-molecule measurements avoid ensemble-averaging the observables, such that heterogeneous subpopulations within an ensemble can be revealed, and transition dynamics between different system states observed without the need of bulk synchronization. In addition to these intrinsic features, however, there are more "perks" that these single-molecule methods can offer. In this review, we will introduce several representatives of single-molecule techniques, specifically those using fluorescence imaging as the detection modality, compare and contrast these techniques with their conventional bulk counterparts, and in doing so parse out the unique advantages that researchers are able to gain by choosing the single-molecule versions of the conventional methods.

*Single-Molecule Real-Time Sequencing Technologies*

Sequencing technologies for nucleic acids have propelled the advancement of biology, biomedical researches, and disease diagnostics. The first-generation sequencing technology, represented by Sanger sequencing, employs the dideoxy chain-termination method, where the incorporation of dideoxynucleoside triphosphates (ddNTPs) terminates DNA polymerase-mediated primer extension reactions, resulting to truncated DNA products that are sequenced via polyacrylamide gel electrophoresis.[23, 24] Despite the high accuracy and a relatively long read length of up to a kilobase, its low-throughput nature limits its application in solving large-scale, omics-level questions. To overcome these constraints, the second-generation sequencing technology (NGS) was invented and rapidly developed during the last 20 years. NGS employs DNA random fragmentation, adapter ligation, and solid-phase amplification to generate high-density monoclonal sequence clusters. Each cluster represents a unique DNA fragment to be sequenced, thus drastically improving the sequencing throughput. In combination with the sequencing-by-synthesis (SBS) chemistry using reversible polymerization terminators, it NGS captures the fluorescence signals generated during the polymerase-driven incorporation of labeled nucleotides and achieves round-by-round base calling from billions of clusters in parallel.[25, 26] Although the template amplification and cluster formation strategy used in NGS significantly increases the signal-to-noise ratio (SNR) and improves the base calling accuracy, this step introduces amplification bias and, more critically, leads to signal decay due to the "phasing" issue (progressive loss of synchronization among molecules within a cluster during successive synthesis cycles). This decay in base-calling accuracy gradually occurs after hundreds of sequencing cycles and significantly compromises the read length.[27-30]

Since the key issue that limits NGS's read length is the progressive loss of synchronization among cluster molecules, conducting sequencing experiments at the single-molecule level can circumvent the need for synchronization and fundamentally addresses the root problem. Building upon this principle, the third-generation sequencing technology (TGS) has emerged and bloomed in the recent years. By directly detecting signals from individual nucleic acid molecules, these methods not only eliminate the systemic biases inherent in NGS due to template amplification, but also enable ultra-long reads (>10 kb) while preserving the base modification information on the original copies of the nucleic acid molecules.[31] For example, Helicos is an early single-molecule sequencing technology based on fluorescence detection and SBS.[32] Similar to NGS, this technology fragments DNA to prepare sequencing libraries and uses polymerases to add fluorescently labeled nucleotides to primers for sequencing.[33,34] However, Helicos directly sequences single DNA molecules instead of clusters, thus avoiding the errors and biases introduced during amplification. One of the common technical difficulties that the fluorescence-based single-molecule sequencing technologies face is the detection of single-molecule signals in the presence of high fluorescence background. Typically, polymerase reactions require micromolar concentrations of fluorescently labeled dNTPs, creating a high fluorescence background that renders conventional single-molecule detection methods inadequate.[35] To address this challenge, Turner *et al.* developed the zero-mode waveguide (ZMW) nanostructures, fabricating ~100 nm diameter aluminum-clad apertures on quartz substrates. In comparison to the relatively large excitation volumes around picoliters for total internal reflection fluorescence (TIRF) and confocal microscopy, these nanostructures in ZMW reduce the excitation volume to the attoliter scale by leveraging the evanescent wave confinement, thus achieving real-time single-molecule observation even at high concentrations of fluorescently labeled dNTPs.[36,37] Using dNTPs with fluorophores covalently attached to the terminal phosphate group, the technology successfully tracked continuous DNA polymerase synthesis and revealed polymerase kinetic states.

In spite of an improved read length, TGS only managed to achieve a modest base calling accuracy in its early stages. To address this issue, Turner *et al.* developed a novel approach termed circular consensus sequencing (CCS), in which the target DNA fragments are ligated to hairpin-shaped sequencing adapters, forming a unique "dumbbell-shaped" circular structure.[38] In this configuration, DNA polymerase performs multiple rounds of rolling-circle amplification around the same DNA molecule.[38] This design ensures each base to be sequenced for multiple times, allowing the generation of high-accuracy consensus sequences from individual reads via redundant observations.[38] To further reduce the error rate, Michael *et al.* optimized CCS by increasing the sequencing coverages, refining the base-calling algorithms, and leveraging an improved polymerase chemistry to enhance the number of passes on the same molecule.[39] These approaches achieved high-fidelity long reads with an accuracy exceeding 99.8%, overcoming the trade-off between read length and accuracy in TGS technologies.[39]

It is worth mentioning that in addition to the fluorescence-based single-molecule sequencing technologies, the nanopore-based sequencing technologies also achieved an excellent base calling accuracy and ultra-long read lengths beyond the limitations of NGS.[40-42] Due to the above-mentioned technical advancements, these single-molecule sequencing technologies have been commercialized, with representative platforms including PacBio's single-molecule real-time (SMRT) sequencing technology and Oxford Nanopore's nanopore sequencing technology (such as the MinION system), significantly driving the paradigm shift in genomics researches. For instance, in the field of genome integrity analysis, Sergey *et al.* combined the PacBio HiFi system with nanopore sequencing to complete telomere-to-telomere (T2T) sequencing of the human genome, filling 8% of the sequence gaps.[43, 44] In the realm of single-cell genomic analysis, Joanna *et al.* developed an automated dMDA (damage-dependent multiple annealing and CRISPR-Cas9-based primer extension) technology for whole-genome amplification of single cells and integrated it with PacBio HiFi sequencing to achieve the comprehensive whole-genome profiling.[45] This method demonstrated an average genomic coverage of up to 40% in single-cell sequencing while enabling the simultaneous detection of genetic heterogeneities, including single-nucleotide variants (SNVs), structural variations (SVs), and tandem repeats.[45] By minimizing the amplification bias and preserving native DNA features, the approach significantly enhanced the resolution and reliability of single-cell genomic characterization.[45] In summary, combining the single-molecule perspective with sequencing technologies not only provides longer reads and more accurate measurement results, but also opens up brand-new analytical dimensions for life science researches.

***Single-Molecule Arrays***

The accurate quantification of ultralow-concentration biomarkers (*e.g.*, proteins, nucleic acids, enzymes, and metabolites) is critical for diverse applications in clinical diagnostics, drug discovery, and environmental monitoring. Conventional methods like PCR and ELISA, in many cases, have proven inadequate for accurate biomarker sensing at low concentrations, as these methods face limitations in sensitivity, matrix interference, and inability to resolve trace-level mixtures,[46] thereby creating an imperative for ultrasensitive detection techniques.[47-51] Furthermore, traditional molecular assays typically offer relative quantification, and require spike-in controls to achieve absolute quantification. One way to circumvent these drawbacks is to introduce digital assays, which presents a transformative alternative by enabling single-molecule counting in partitioned samples.[52] Taking advantage of eliminating normalization dependencies and realizing absolute quantification without external calibration, this single-molecule array (Simoa) technology stands out as an ultrasensitive digital platform capable of detecting target molecules—primarily proteins but also nucleic acids and other biomolecules—with single-molecule sensitivity. For example, Simoa platforms, such as digital enzyme-linked immunosorbent assay (dELISA), have enhanced measurement sensitivities for various proteins by up to 1000-fold compared to traditional ELISA.[50, 53, 54]

In the Simoa prototype system, target analytes are captured on biofunctionalized paramagnetic beads (MBs), with a large excess of beads ensuring Poisson-distributed single-molecule binding (0 or 1 molecule per bead).[55] The fluorophore-labeled microbeads are then isolated in an array of femtoliter-sized microwells designed for single-bead occupancy. After sealing the Simoa substrate with oil to facilitate reagent reactions, enzyme-catalyzed fluorescent products remain confined within individual micron-sized compartments. Only wells containing one target molecule generate discrete, localized fluorescence outputs that can be easily detected by an optical imaging system.[56, 57] This approach therefore enables background-free detection with femtomolar to attomolar sensitivity through digital counting statistics.

Assays particularly for tracing enzymatic activity play a pivotal role in clinical diagnostics, functional proteomics, and drug discovery.[58] However, conventional enzyme assays are only intended for measuring analyte concentrations in a homogeneous bulk solution, but not for real-time monitoring of heterogeneous, dynamic enzymatic events. Walt *et al.* developed a method for absolute quantification of enzymatic activity using Simoa (eSimoa).[59] In eSimoa, substrate-conjugated MBs are incubated with the target enzyme in the presence of cofactors. The enzyme converts the substrate into a fluorescently detectable product on the bead surface for Simoa analysis. The eSimoa method has demonstrated exceptional sensitivity in detecting enzymes such as protein kinases, telomerase, and histone methyltransferase. Additionally, through enzyme inhibition assays and binding affinity calculations, eSimoa can evaluate tight-binding inhibitors, successfully determining inhibition constants for drugs such as Bosutinib and Dasatinib. Finally, the ability to measure enzymatic activity at the single cell level enables the in-depth investigation of cellular heterogeneity in drug response, which has great implications to the treatment of cancers.

While Simoa achieves subfemtomolar detection limits and remains the gold standard for ultrasensitive protein analysis, its sensitivity is fundamentally constrained by low sampling efficiencies.[60] To address this limitation, bead encapsulation in water-in-oil droplets has emerged as a promising strategy to enhance sampling efficiency in digital bioassays. Recent advances in digital droplet-based immunoassays demonstrate bead loading efficiency of up to 60%, achieving sensitivities that rival or even surpass (by up to an order of magnitude) conventional Simoa technology.[54, 61] However, their integration into point-of-care (POC) systems is complicated by the requirement for precisely controlled droplet generation. Besides, a major challenge for POC implementation lies in the throughput limitation caused by the requirement to image numerous bead-free droplets. Walt *et al.* addressed these limitations by developing dropcast single-molecule assays (dSimoa), an innovative approach that combines on-bead signal generation with bead dropcasting into monolayer films for single-molecule counting.[62] This platform offers several key advantages: (i) By localizing a nondiffusible fluorescent signal to each target-carrying bead, it eliminates the need for bead loading into microwells or droplets for signal compartmentalization; (ii) It enables analysis of significantly more beads, thereby improving sampling efficiency and enhancing sensitivity; (iii) The template-free readout only requires a basic optical setup for signal transduction, thus offering improved cost-effectiveness for POC integration. These advancements allow the dSimoa platform to achieve attomolar detection limits, with an up to 25-fold sensitivity improvement over traditional Simoa technology, and represent the current state of the art for ultrasensitive protein detection.

The Simoa platform has also shown promise for nucleic acid detection, including microRNAs (miRNAs), which serve as valuable noninvasive biomarkers despite their extremely low abundance in urine (2-4 orders of magnitude lower than in blood).[63, 64] While exponential isothermal amplification (EXPAR) enables rapid miRNA quantification, its multiplexing capability is limited by spurious amplification, poor sensitivity and reproducibility.[65, 66] Li *et al.* developed a droplet-based digital EXPAR (ddEXPAR) platform that combines immunomagnetic exosome capture with compartmentalized amplification, enabling femtomolar-sensitive quantification of miRNAs in microwells.[67] This approach generates background-free, heterogeneous fluorescence profiles that differs from the conventional homogeneous solution-based assays for nucleic acid amplification, and distinguished healthy individuals from primary urethral carcinoma patients and further classified disease subtypes. Compared to traditional solution-based EXPAR, this digital assay significantly reduces non-specific amplification common occurred in bulk solution and improves sensitivity by over 2 orders of magnitude. While there are other methods for digital analysis of nucleic acids using quantitative polymerase chain reaction (qPCR)[68, 69] or isothermal amplification strategies,[57, 70, 71] the exponential nature of such amplifications renders it prone to leaky reactions and inevitably induce spurious, non-specific amplification reactions. One solution is to embed amplification reagents in a nanoporous hydrogel matrix that provides ideal permeability to allow free diffusion of nucleic acids and easy removal of unreacted primers.[72]

In summary, Simoa technology demonstrates pioneering advancement in bioanalytical detection by achieving absolute quantification at attomolar sensitivity through its single-molecule counting methodology. Concurrent technological innovations have progressively improved the sensitivity, simplicity and adaptability of such platforms: eSimoa enables ultrasensitive quantification of enzymatic activity; the streamlined dSimoa platform is able to reach attomolar sensitivity for common protein determination; ddEXPAR as a useful complement to digital PCR significantly enhances specificity for multiplexed miRNA analysis in human bodily fluids. These novelties collectively improve the clinical precision of biomarker analysis and optimize early disease diagnosis. In addition, Simoa technology exhibits superior biocompatibility with complex biological matrices and enables robust multiplexed detection, positioning it as a viable standardized platform for translational medicine applications. With continued development, Simoa is poised to play an increasingly critical role in early disease diagnosis and post-therapeutic recurrence monitoring.

### *Single-Molecule Pull-Down Assays*

Protein-protein interactions (PPIs) are crucial in understanding many biological processes and have long been a central topic in biological mechanism studies.[73-75] To study the direct or indirect PPIs, pull-down assays are the gold standard widely used in the field.[76] In pull-down assays, bait proteins carrying affinity tags specifically bind to the antibodies or substrates immobilized on beads, capturing the bait proteins together with their interacting partners. With many types of orthogonal affinity tags and their corresponding beads (*e.g.*, GST, FLAG, HA tags) developed to facilitate the applicability,[77] complexes-of-interest are routinely pulled down, and the captured proteins easily eluted for downstream analysis with Western blot[78, 79] or mass spectrometry methods.[80] Despite its simplicity and versatility, pull-down experiments face certain limitations, such as the inability to distinguish mixed complex subpopulations and to determine the complex stoichiometry, the difficulty in capturing weak or transient interactions,[81, 82] and the incompatibility with large-scale PPI screening.

To overcome these drawbacks, the Ha group developed an ingenious yet simple method for studying cellular protein complexes at the single-molecule level. In this method, named the single-molecule pull-down assays (SiMPull),[83, 84] physiological macromolecular complexes are directly pulled down from cell or tissue extracts onto the imaging surface on a single-molecule fluorescence microscope, which is pre-coated with antibodies that specifically recognize the tag-of-choice. Typically, the antibody coating is achieved using the PEG-biotin-streptavidin system, where the surface is functionalized with a PEG-based polymer layer (*e.g.*, linear or branched PEG,[85, 86] or PEG-containing detergents[87]) that also sparsely carries biotin groups for attachment; on top of this layer, streptavidin (or NeutrAvidin) acts as "glue" to tether the biotinylated antibodies. Besides, the PEG coating also serves the purpose of passivation to prevent the adsorption and interference of non-specific proteins. Flow channels are constructed on the imaging surface to facilitate buffer exchange. Cell lysates that contain the fluorescently labeled bait proteins are then introduced on the imaging surface, leading to the co-immunoprecipitation of the interacting partners, which are also labeled with fluorophores for detection. After washing out the unbound components, the target complexes are observed and analyzed through the TIRF microscopy, generating single-molecule signals with a high SNR.[83, 84]

As an excellent example for the effectiveness of SiMPull, Jain *et al.* studied the mTOR kinase-related complexes and confirmed that both mTORC1 and mTORC2 are dimeric in composition. Furthermore, the study revealed that although mTORC1 and mTORC2 are dimeric overall, their individual components (such as mTOR, raptor, rictor, *etc.*) primarily exist as monomers when expressed separately. This finding indicates that the dimerization of mTORC1 and mTORC2 is not driven by interactions between any single component, but rather is the result of multiple subunits cooperating to form a composite surface. The authors then investigated the effects of different physiological conditions (such as nutrient deprivation, energy stress) and pharmaceutical treatments (such as rapamycin) on the mTOR complex assembly and stoichiometry. The results showed that most of the conditions and treatments do not lead to the alteration of the dimeric nature of the mTOR complexes; only rapamycin partially disrupts mTORC1, leading to transient monomeric mTOR intermediates, while mTORC2 remains unaffected.[88]

Compared to the traditional pull-down techniques, performing pull-down experiments at the single-molecule level has several obvious advantages. 1) To measure the copy number of fluorescently labeled molecules in the complex, SiMPull experiments record the photobleaching trajectories of the fluorescent labels (*i.e.*, the number of sudden step-wise drops in the fluorescence intensity). After accounting for the maturation and labeling efficiency of the fluorescent labels, the stoichiometry of the target proteins in the complex can be inferred based on the distribution of photobleaching steps. 2) SiMPull can distinguish complex subpopulations in a heterogeneous mixture that are often obscured under ensemble measurements. For example, a 1:1 mixture of two complexes, $AB_3$ and $AB$, could be mistakenly determined as $AB_2$ in stoichiometry if not measured at the single-molecule level. 3) SiMPull allows for the real-time observation of dynamic changes in protein complexes and the determination of protein-protein interaction kinetics. 4) SiMPull has extremely high sensitivity, capable of detecting and quantifying low-abundance complexes. 5) SiMPull requires a significantly lower sample volume than bulk methods, making it possible to study cells or tissues that are difficult to culture in large quantities. 6) Finally, the SiMPull experimental procedure can be integrated with high-throughput imaging platforms, making it compatible for large-scale screening and analysis.

*Fluorescence Intensity Shift Assays*

Electrophoretic mobility shift assay (EMSA) is a widely used technique in molecular biology to study nucleic acid-binding proteins, such as transcription factors, chromatin remodelers, and ribonucleoprotein complexes.[89-93] The basic principle of EMSA involves the electrophoresis separation of protein-DNA complexes from free DNA under non-denaturing conditions. The DNA band on the gel is shifted when it is bound to a protein, hence the name "mobility shift".[94] Based on the protein and nucleic acid concentrations used, reaction times and mobility shift patterns observed, one can deduce a rich set of information from EMSA, such as the binding affinity, kinetics, and stoichiometry.[95, 96] Moreover, EMSA has been used in combination with microfluidic[97] and nanomaterial-enhanced detection[98], and applied in disease-related researches[99]. Despite its widespread utility in detecting protein-nucleic acid interactions, EMSA has several critical limitations. Most notably, due to a relatively long separation time on gels, transient biomolecular complexes can dissociate during electrophoresis. Therefore, EMSA suffers a poor sensitivity for studying such interactions, even when radiolabeled nucleic acids are used for detection.[100] In addition, once complexes are separated on gels, it is difficult to sequentially add or remove reagents (*e.g.*, another protein or nucleic acid) to study dissociation and the sequential multi-component interactions.

Inspired by the development of SiMPull, Cai *et al.* developed a fluorescence intensity shift assay (FISA) which effectively overcomes the aforementioned limitations of EMSA.[101] Based on the protein-induced fluorescence enhancement (PIFE) effect,[102, 103] FISA detects the nucleic acid-protein interactions by positioning a fluorophore (such as Cy3 and Cy5) in the vicinity of the protein binding sites on the nucleic acids and monitoring the fluorescence intensity shifts due to protein binding.[103, 104] To achieve a strong PIFE effect while minimizing interference with the molecular interactions, the choice and labeling positions of the fluorophore are empirically tested and carefully validated during pilot experiments.[105] As an example, using single-molecule PIFE-FISA, Cai *et al.* investigated the binding kinetics between the bacterial heat-shock regulator CtsR and its target DNA in real time, as well as the modulation mechanism of the CtsR-DNA complex by the arginine kinase McsB.[101, 106] The quantitative FISA measurements showed that CtsR binds rapidly and stably to the target DNA with a $K_D$ and Hill coefficient of 12 nM and 2.1, respectively. Besides, it is revealed that McsB transiently interacts with the CtsR-DNA complex, with the $k_{on}$ and $k_{off}$ rates of 0.75 $\mu M^{-1}$ $s^{-1}$ and 0.34 $s^{-1}$, respectively. The McsB treatments render the CtsR-DNA complex more sensitive to elevated temperatures, as evidenced by the fluorescence intensity down-shift in FISA indicating rapid CtsR dissociation upon heated buffer washes. This kinetic information obtained by FISA, in combination with mass spectrometry and molecular simulation results, together suggest a new hierarchical phosphorylation mechanism of McsB, which highlights the functional importance of periphery arginine residues on CtsR regulation.

Several clear advantages of FISA stand out in comparison to EMSA. 1) FISA provides dynamic information on weak and transient interactions that would likely be lost during gel electrophoresis (*e.g.*, the interaction between McsB and the CtsR-DNA complex with a $K_D$ around 400 nM). 2) FISA allows a low sample volume and easy buffer exchange thanks to the incorporation with flow channel configuration. This benefit is particularly evident when FISA is used to determine the temperature sensitivity of the CtsR-DNA complex, where a rapid introduction of heated buffers and quick washout of dissociated proteins is necessary. 3) Compared to other single-molecule methods that study protein-nucleic acid interactions (*e.g.*, colocalization[107] and FRET[85, 108]), PIFE-FISA bypasses the need of labeling proteins, which not only avoids the non-specific binding of fluorescently labeled proteins, but also overcomes the "concentration barrier" of detecting single-molecule signals in the presence of high concentrations of labeled species in solution.[35] As a result, the researchers can readily use a protein concentration of 200 nM to observe the transient McsB binding events to the CtsR-DNA complex.

***Single-molecule Electrochemiluminescence***

Electrochemiluminescence (ECL) is an optoelectronic phenomenon where luminophores are electrochemically excited on the electrodes at controlled potentials, which leads to light emission facilitated by co-reactants.[109, 110] Since ECL is induced by electrochemical reactions rather than external light, this technique exhibits virtually no background signal, therefore enabling ultrasensitive optical signal readout and affording an excellent SNR.[111, 112] Thanks to these desirable characteristics, ECL has been widely employed as a readout method in a broad set of high-sensitivity bioimaging and bioanalytic diagnostic techniques, including ECL-based antigen imaging,[113] viral protein identification,[114] and ultrasensitive immunoassays.[115, 116] These methods are distinguished by their simplicity, low detection limits, and good reproducibility, and have been extensively integrated into commercial automated analysis platforms.[117] However, conventional ECL measurements yield only averaged signals from an ensemble of luminophores, posing significant challenges in observing individual molecular differences, transient kinetic behaviors, and spatial heterogeneity during the reaction process.[118, 119] This averaging effect tends to obscure the intrinsic complexity of the reaction systems, thereby limiting the understanding of the underlying microscopic mechanisms.[120, 121] Although the development of ECL imaging has improved spatial resolution to some extent, there remains a significant gap in the ECL field compared to that achieved by single-molecule super-resolution fluorescence imaging techniques.[122]

Detecting single ECL reactions with high spatial resolution has long been a goal in the field of single-molecule imaging.[123-125] However, unlike most single-molecule fluorescence techniques, where the SNR of detecting a single fluorophore can be boosted by the hundreds to thousands of photons the fluorophore generates, it is extremely difficult to detect a single ECL luminophore because only a few photons are generated each time the luminophore is excited.[126, 127] To tackle these obstacles, the Feng group has pioneered the development of a highly sensitive ECL microscopy platform, capable of achieving synchronized electrochemical control with wide-field optical detection.[128] Specifically, this platform integrates a standard three-electrode configuration with an inverted optical microscope and utilizes a custom-designed transparent indium tin oxide (ITO) electrode, which ensures the electrochemical reactivity and optical transparency both required in this experiment. By applying precisely controlled potentials and diluting the reactant concentrations to reduce the molecular collision frequency, single-molecule events of ECL reactions have been successfully isolated in both space and time. Facilitated by a high-numerical aperture (NA) objective lens and an electron-multiplying charge-coupled device (EMCCD) detector, the platform achieves the high sensitivity detection of single-photon emission events characteristic of single ECL reactions in solution.

Additionally, Feng *et al.* further exploited the stochastic spatiotemporal distribution of single-molecule ECL signals and applied an image reconstruction strategy to achieve super-resolution imaging. Although individual ECL events generate only a few photons, constraining the precision of conventional single-molecule localization algorithms, repeated emissions at the same reaction site can form multiple localization clusters in adjacent regions, thereby substantially enhancing the spatial resolution of image reconstruction. As a result, the platform achieves a spatial resolution of approximately 22 nm in the $Ru(bpy)_3^{2+}$/TPrA system and was successfully applied to the dynamic super-resolution imaging of live cells, thus establishing a label-free, laser-free, and real-time imaging methodology based on the single-molecule ECL microscopy.[128] Specifically, during live-cell imaging, the system demonstrated a spatial and temporal resolution of 150 nm and 12 seconds, respectively, which are comparable to those achieved by optimized super-resolution fluorescence microscopy. Furthermore, employing the single-molecule ECL microscopy, researchers achieved super-resolution imaging of $Ru(bpy)_3^{2+}$-mediated reactions on gold (Au) electrode surfaces and investigated the underlying mechanisms of electrocatalytic reactions. This research revealed that structural reorganization of the catalyst during reactions plays a pivotal role in determining its activity, providing a new insight into the dynamic relationship between surface structures and catalytic performance.[129]

Compared to the conventional ECL techniques, single-molecule ECL imaging overcomes the intrinsic limitations of ensemble-averaged measurements, enabling ultrasensitive and super-resolution detection of individual chemical reaction events.[130,131] This method offers direct molecular-level insights into catalytic reactions,[132,133] molecular recognition,[134] and electron transfer processes,[135,136] revealing critical details that are often obscured in traditional ensemble measurements. With broad application prospects in areas such as single-cell analysis,[137,138] tracking membrane protein dynamics,[139] and investigating interfacial reactions in materials chemistry,[140] single-molecule ECL lays a solid foundation for the development of next-generation functional imaging technologies, and opens new frontiers for understanding chemical processes at the molecular level.

*Other Single-Molecule Versions of Ensemble Biochemical Assays*

The strategy of augmenting ensemble biochemical assays with single-molecule capability is definitely not limited to the several examples discussed above. For example, to measure the kinetics and equilibrium parameters of biomolecular interactions, the Tao group developed the plasmonic scattering microscopy (PSM),[141] which can be viewed as the single-molecule counterpart of the surface plasmon resonance (SPR) method and is able to achieve digital kinetic analysis of individual protein molecules while obtaining the size information of the binding proteins, as demonstrated via detecting single molecules of human immunoglobulin M (IgM) and immunoglobulin A (IgA). In addition, a super-resolution single-molecule fluorescence binding assay has also been developed by the Ha group, where the accuracy of multi-color colocalization between ligands and receptors is enhanced via the application of the centroid localization algorithm.[142] This innovation effectively corrects for common artifacts including nearby non-specific binding events and optically overlapping binding sites at the diffraction-limited resolution. Once again, various advantages have been amply demonstrated in these examples.

*Perspectives*

In summary, through the examples discussed, we have seen many attractive features of the single-molecule tools and methods, from negating the need for bulk synchronization to observe state transitions in SMRT sequencing, to digital counting and absolute quantification of biomolecules in Simoa, from unmasking molecular heterogeneity and determining stoichiometry in SiMPull, to easy integration with flow channel configuration and *in situ* imaging in FISA, and finally to the capability of spatial super-resolution in single-molecule kinetic assays and ECL. The superior performance of these methods, such as ultra-long reads, richer information and fewer artifacts, as well as the ability to work with weak or transient interactions, are arguably the direct results from these unique technical advantages.

Can more ensemble biochemical assays be converted into their single-molecule versions? The answer is probably yes. Here we would like to propose a new single-molecule assay, named "single-molecule polysome profiling". The conventional polysome profiling technique, which sediments and separates polysomes in a density gradient under ultracentrifugation conditions, is the standard (and perhaps the only) method available to determine the overall distribution of translating ribosomes in cells. Though being broadly applied, this technique barely resolves heavy polysomes containing more than ten ribosomes, and only provides a semi-quantitative estimate of the amounts of different polysomes, not mentioning the lengthy operation it demands on an ultracentrifuge. Inspired by the subunit counting method in SiMPull, it is reasonable to assume that one can probably count the number of ribosomes in a polysome complex, if each ribosome is fluorescently labeled and the polysome complex immobilized on the imaging surface. Since the number of resolvable fluorescence intensity levels are readily around three or four in FISA,[101] and can reach up to 20 as demonstrated by the Ha group,[143] it is foreseeable that the single-molecule polysome profiling assay could not only detect larger polysome complexes in a more quantitative way, but also significantly reduce the sample amount and assay time.

While assays like ELISA, pull-down experiments, and EMSA are routinely carried out in many biochemistry labs, each assay requires its own set of apparatus, and has a different operation protocol with specific steps that are not transferrable among assays; some of the instruments required for these assays (*e.g.*, a plate reader) can be rather expensive. We cannot help but wonder what would happen if the imaging-based single-molecule versions of many common assays were fully developed and adopted as the "bread-and-butter" tools in labs. Not only one could access all the benefits of single-molecule experiments discussed above, but also the number of different lab instruments would drastically decrease; perhaps one carefully designed multi-purpose microscope could suffice. As a result, all these experiments could be performed on the same microscope platform and the difference in operating different assays could be as trivial as just adding different buffers in the flow channels. These efforts would inevitably promote a wider accessibility and applicability of new biochemical methods and tools.

**Acknowledgments**
This work was supported by the Fundamental Research Funds for the Central Universities 2024300410 (to B.H.), State Key Laboratory of Analytical Chemistry for Life Science, Nanjing University 5431ZZXM2403 (to B.H.), the National Natural Science Foundation of China No. 22404079 (to B.H.), and the National Natural Science Foundation of China No. 22076125 (to Z.L.).

**Author contributions**

**H.T., S.L., C.K., X.W., X.Z., K.L., G.D.**: Writing-Original Draft; **Z.L.**: Writing-Original Draft, Supervision, Funding acquisition; **B.H.**: Conceptualization, Writing-Original Draft, Supervision, Funding acquisition

**Conflict of interests**

The authors declare no competing financial interest.